# Relativistic poloidal non-uniformity of the radial electric field in tokamak plasmas.

Romannikov A.N. *, Khvostenko P.P. **.

* - JSC "SSC RF TRINITI", Romannikov@rtiniti.ru

**- SIC "Kurchatov Institute"

**ABSTRACT:**

The paper presents a relativistic mechanism that can create a poloidal non-uniformity of a radial electric field on a magnetic surface in a tokamak plasma. This mechanism is associated with a significant current velocity of electrons in plasma for many experimental modes of tokamak operations. It is important that electric fields arise without the traditional separation of charges in plasma. The presented mechanism can explain the previously experimentally measured poloidal non-uniformity of the $C^{+6}$ carbon ion toroidal rotation velocity in the TORE-SUPRA tokamak plasma [1].

## Introduction

Traditionally, relativistic effects are neglected when describing plasma, at least in small and medium tokamaks with average densities (several units per $10^{13}$ cm-3) and temperatures (up to 1 keV). The reason is quite clear – it is a small ratio of particle group velocities in plasma to the speed of the light ($\frac{V_{gv}^2}{c^2} < 10^{-5}$). The accuracy of measuring plasma parameters for diagnostic complexes of real tokamaks practically does not allow measuring such relativistic effects.

The radial electric field $E_r(r)$ in a cylindrical tokamak plasma is studied and calculated within the framework of the radial balance of forces on magnetic surfaces. This makes it possible to obtain a relation connecting the velocity of ion toroidal and poloidal rotation, the radial electric field and the so-called diamagnetic term:

$$E_r(r) \cong \frac{V_\varphi(r) \cdot B_\vartheta(r)}{c} + \frac{1}{Z|e| \cdot n_i(r)} \cdot \frac{dp_i(r)}{dr} - \frac{V_\vartheta(r) \cdot B_\varphi(r)}{c} \tag{1}$$

where $c$ is the speed of light, $Z|e|$ is the charge of ions, $n_i(r)$ and $p_i(r)$ is the density and pressure of plasma ions, $V_\varphi(r)$ is the ion toroidal rotation velocity, $V_\vartheta(r)$ is the ion poloidal rotation velocity, $B_\vartheta(r)$ is the poloidal magnetic field, $r$ is the radius of the magnetic surface, $\vartheta$ is the poloidal angle. The poloidal rotation velocity in formula (1) can be calculated from the "neoclassical" theory. It give us $V_\varphi(r) \cong c \dfrac{k}{Z|e|B_\vartheta(r)} \dfrac{\partial T_i(r)}{\partial r}$; where $T_i(r)$ is the plasma ion temperature температура ионов плазмы, $k$ varies from 1.17 for a weakly collisional "banana" regime to -2.2 for a collisional plasma [2].

Unfortunately, formula (1) is not enough to calculate $V_\varphi(r)$ and $E_r(r)$ in axisymmetric tokamak configurations. It would be possible to use the ambipolary equation for radial flows, but automatic ambipolarity occurs in the configuration under consideration (see, for example, [3]). This leads to the fact that $E_r(r)$ (or $V_\varphi(r)$) is a free parameter for such plasma configurations. It is possible to introduce additional equations in some experimental situations. This may be an additional ion flux in a tokamak with significant toroidal magnetic field ripples [4, 5], or additional fast ion flows in the case of ion-cyclotron heating or in the case of using heating neutral beams, [6, 7, 8], and etc.

In all the presented approaches $E_r(r)$ on the magnetic surface is considered practically non-uniformity on a poloidal direction, and the plasma on the magnetic surface rotates on the toroidal direction as a solid body. It is usually measurements of $E_r(r)$ (or $V_\varphi(r)$) poloidal dependence on one magnetic surface not carried out in tokamaks. It can be noted the experiment for measuring of $E_r(r)$ poloidal dependence in stellarator plasmas [9]. Within the limits of measurement accuracy, the result was predictable – there was no dependence.

It is also possible to note experiments on measuring the poloidal non-uniformity of the toroidal rotation velocity of carbon ions $C^{+6}$ in TORE-SUPRA tokamak plasmas [1], in which the plasma on the magnetic surface rotated in the toroidal direction with a significant difference from the rotation as a solid body.

The method traditionally used in tokamaks for measuring plasma rotation velocities and further calculation is the charge exchange recombination spectroscopy (CXRS). Unfortunately, all measurements are carried out most often from the weak field from the periphery to the center of the plasma, which does not allow measuring the poloidal dependence, for example $V_\varphi(r)$, at least at several points on the same magnetic surface.

A relativistic mechanism for the possible occurrence of poloidal non-uniformity $E_r(r)$ on a magnetic surface is presented in the article. It is based on the relativistic expression of the electric field in the laboratory coordinate frame for the case of a rapidly moving electron. It is shown that the magnitude of this non-uniformity is quite measurable. It is important, there is no traditional separation of charges in the plasma for the appearance of an electric field. The presented mechanism can explain the experimentally measured poloidal non-uniformity of the $C^{+6}$ toroidal rotation velocity in the TORE-SUPRA tokamak plasma [1].

$E_r(r)$ has been the subject of research on many tokamaks for a long time (see, for example, [10-13]). Special profiles of the radial electric field, as well as the associated toroidal and poloidal plasma rotation profiles, can create the appearance of improved modes (H-modes and etc.) in tokamaks. The mechanisms of occurrence of such regimes, which are hoped for in experiments on ITER and the next generations of tokamaks, remain currently not fully understood. The mechanism presented in this paper can partially explain the nuances of a radial electric field $E_r(r)$ creation process in the tokamak plasma.

## RELATIVISTIC DEPENDENCE OF Er(r) ON PLASMA CURRENT

Let's consider the initially simple model presented in FIG. 1. Plasma consists of electrons that move with current velocity $V_I$ and practically nonmoving ions. The plasma is partially enclosed in an electrically grounded metal chamber. It is necessary to calculate the electric field $E_r$ at a distance **a** at the point with an asterisk. **2L₀** is the approximate length of the plasma that can participate in the creation of the $E_r$ (not shielded by a grounded chamber).

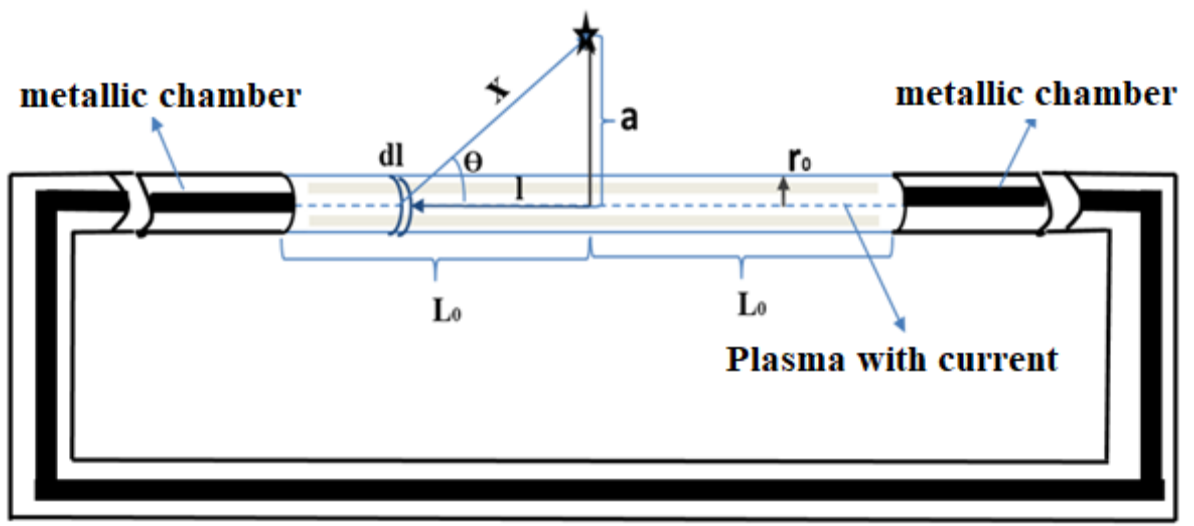

FIG.1. Plasma with current in a metallic chamber.

An electron with a charge **e** moving at such a velocity **V_I**, say on the plasma axis in the region dl, creates an electric field [14] at a point at a distance **a** in the direction of a small radius **r₀**, equal

to:

$$\frac{e}{x^2} \cdot \frac{(1 - \frac{V_I^2}{c^2}) \cdot \sin\theta}{(1 - \frac{V_I^2}{c^2} \cdot \sin^2\theta)^{3/2}} \qquad (2).$$

The electric field of a rapidly moving electron turns into a "pancake". In the direction perpendicular to the motion, the electric field increases with respect to the spherical Coulomb field, and in the direction of motion it decreases.

Assuming that the density of electrons **n** is equal to the density of ions in plasma and taking into account the symmetry of the model, it is possible to calculate the radial electric field created at the point with the asterisk by electrons and ions. We have for electrons with current velocity:

$$E_r^e(a)_- \cong \frac{e \cdot n \cdot \pi \cdot r_o^2 \cdot dl}{x^2} \cdot \frac{(1 - \frac{V_I^2}{c^2}) \cdot \sin\theta}{(1 - \frac{V_I^2}{c^2} \cdot \sin^2\theta)^{3/2}}$$

$$E_r^e(a)_- \cong 2 \cdot \int_0^{L_0} \frac{e \cdot n \cdot \pi \cdot r_o^2}{l^2 + a^2} \cdot \frac{(1 - \frac{V_I^2}{c^2}) \cdot \frac{a}{\sqrt{l^2 + a^2}}}{(1 - \frac{V_I^2}{c^2} \cdot \frac{a^2}{l^2 + a^2})^{3/2}} \cdot dl =$$

$$= \frac{2 \cdot e \cdot n \cdot \pi \cdot r_o^2}{a} \cdot \frac{1}{(1 + \frac{a^2}{L_0^2})^{1/2}} \cdot \frac{1}{\sqrt{1 - \frac{a^2}{L_0^2 + a^2} \cdot \frac{V_I^2}{c^2}}} \qquad (3).$$

For nonmoving ions, the last multiplier is missing in the equation (3) and the charge sign is positive. Thus, the total negative radial electric field associated with the relativistic factor $\frac{V_I^2}{c^2}$ arises at the point with the asterisk. So, we have:

$$E_r(a) \cong \frac{e \cdot n \cdot \pi \cdot r_o^2 \cdot}{a} \cdot \frac{a^2}{L_0^2} \cdot \frac{1}{(1 + \frac{a^2}{L_0^2})^{3/2}} \cdot \frac{V_I^2}{c^2} \qquad (4).$$

If we calculate the electric fields at a distance **a** near the edge of the metal chambers, we can show that the radial component in this area becomes close to zero, and the longitudinal component reaches a maximum. In this case, the field is directed from the edges of the plasma towards the point with an asterisk in FIG.2. Thus, we have a quadrupole electric field.

But, $e \cdot n \cdot \pi \cdot r_o^2 \cdot V_I$ is the plasma current **I$_p$**. We do not consider the possible nuances of the connection of the plasma current with the electron function distribution in the framework of kinetic theory. Therefore, $E_r^e(a)$ can be expressed in terms of the total current:

$$E_r(a) \cong \frac{I_p^2}{e \cdot n \cdot \pi \cdot r_o^2 \cdot a} \cdot \frac{a^2}{L_0^2} \cdot \frac{1}{(1 + \frac{a^2}{L_0^2})^{3/2}} \cdot \frac{1}{c^2} \qquad (5).$$

It may seem that there is a problem with the Gauss theorem - $\oint_S (\vec{E} \cdot d\vec{S}) = 4\pi \sum_i q_i$, where $\vec{E}$ is the electric field vector on an element of a closed surface $d\vec{S}$, $q_i$ are the charges inside the surface. The relativistic equation (2) of Coulomb's law does not affect the Gauss theorem. At points on a closed surface, the electric field can be any, depending on the distribution of charges inside. It is essential that $\oint_S (\vec{E} \cdot d\vec{S}) = 0$, and this holds for every selected closed surface in the model under consideration.

The question arises, what will happen to such a relativistic $E_r(r)$ if the plasma with a current and the chamber will be closed into a toroid, as in a tokamak. The average plasma current velocity V$_I$ in typical ohmic discharges for almost all small and medium tokamaks is about 300 km/s. Indeed, on the T-11M with the parameters: **R(см)/r$_0$(см)~70/** (hereinafter, **r$_0$** is a small tokamak radius), with a typical plasma current Ip ~50-60 kA and an average plasma density **<n$_e$>** ~$10^{13}$ cm$^{-3}$; - the average plasma current velocity is ~ 300 km/s.

On the tokamak T-10 ($\bar{I}_p \sim 350$ кА, **R(см)/ r$_0$ (см)**~150/30, **<n$_e$>** ~2 ×10$^{13}$ cm$^{-3}$) - V$_I$ ~350 km/s. For tokamak T-15MD (Ip ~2 MA, **R(см)/ r$_0$ (см)**~148/67, elongation 1.9, **<n$_e$>** ~5 ×10$^{13}$ cm$^{-3}$) - V$_I$ ~320 km/s.

The metal chamber of the tokamak in real experiments is always electrically grounded. The resistance of the camera to the ground for various tokamaks can range from tens of kOhm to 10 MOhm. At the same time, the RC of the camera for various tokamaks can vary from several tens of microseconds to several milliseconds. Thus, it is necessary to take into account the effect of shielding with a grounded camera on the formation process of $E_r(r)$ within the framework of the proposed approach.

Consider a simple model for two points with respect to a toroidal plasma, as shown in FIG.2. We initially assume that the camera is not electrically grounded and does not influence on $E_r(a)$ in the points under study.

**(a)**

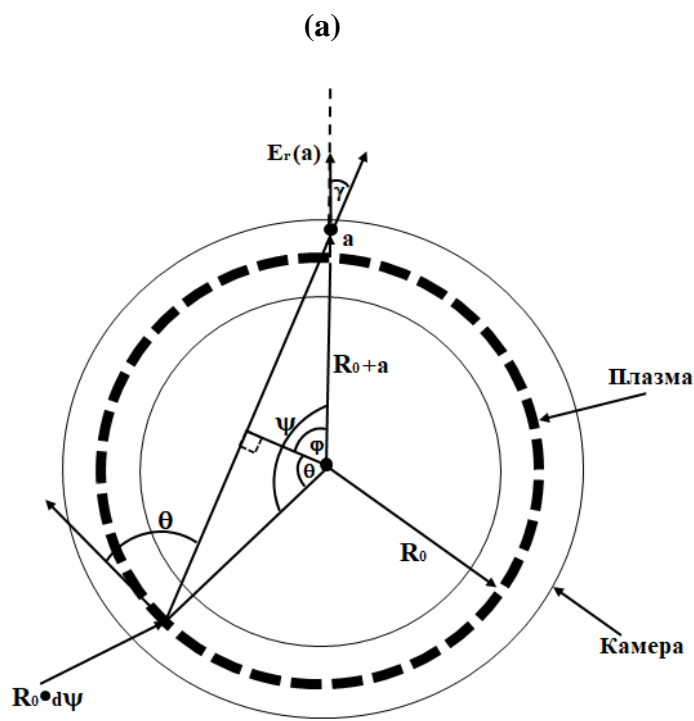

**(b)**

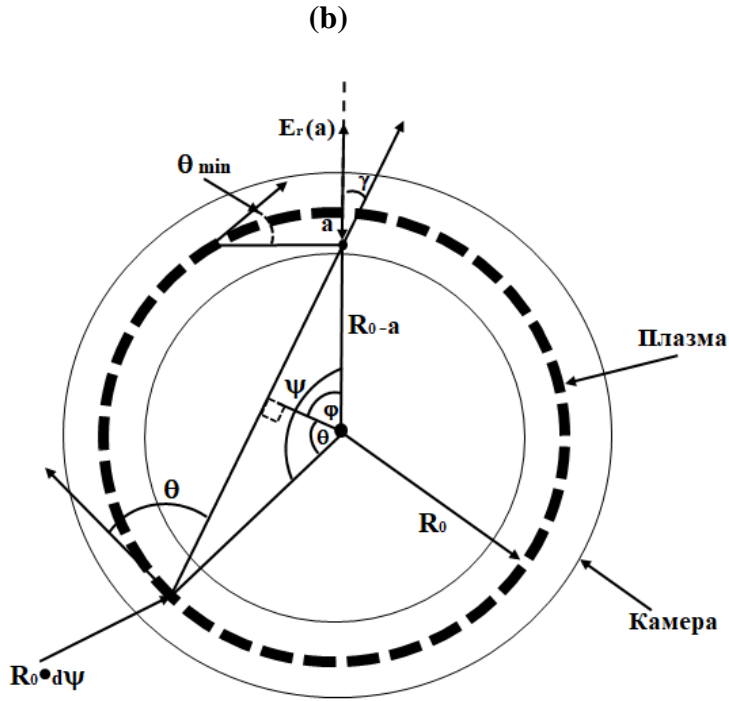

FIG.2. Geometric model for $E_r(a)$ calculation at points $R_0 + a$, **(a)** and $R_0 - a$, **(b)**.

In a similar geometry, a term associated with the centrifugal acceleration of a ring with a plasma current appears in equation (2) [14]. But, as noted in [14], this term becomes important for $a > R_0$. Therefore, in our calculations, with $a$ noticeably less $R_0$, it can be neglected. The calculation of the radial electric field is symmetric when changing $\psi(0 \rightarrow \pi) \Leftrightarrow \psi(\pi \rightarrow 2\pi)$.

For simplicity, we assume that the plasma with current occupies the region with $r_0 << a$.

Для простоты считаем, что плазма с током находится в области с $r_0 << a$.

Рассчитаем электрическое поле $E_r(a)$ для случая а), РИС.2, для электронов с токовой скоростью $V_I$.

Calculate the electric field $E_r(a)$ for case a), FIG.2, for electrons with current velocity

$V_I$.

$$dE_r(a)_{-} \cong \frac{-|e| \cdot \overline{n} \cdot \pi \cdot r_0^2 \cdot R_0 \, d\psi}{x^2} \cdot \frac{(1 - \frac{V_I^2}{c^2}) \cdot \sin\gamma}{(1 - \frac{V_I^2}{c^2} \cdot \sin^2\theta)^{3/2}} \cong$$

$$\cong \frac{-|e| \cdot \overline{n} \cdot \pi \cdot r_0^2 \cdot R_0 \, d\theta \cdot (1 + \frac{\frac{R_0}{R_0 + a} \cdot \sin\theta}{\sqrt{1 - (\frac{R_0}{R_0 + a})^2 \cdot \cos^2\theta}})}{(R_0 \sin\theta + (R_0 + a)\sqrt{1 - (\frac{R_0}{R_0 + a})^2 \cdot \cos^2\theta})^2} \times \frac{(1 - \frac{V_I^2}{c^2}) \cdot \sqrt{1 - (\frac{R_0}{R_0 + a})^2 \cdot \cos^2\theta}}{(1 - \frac{V_I^2}{c^2} \cdot \sin^2\theta)^{3/2}} \cong$$

$$\cong \frac{-|e| \cdot \overline{n} \cdot \pi \cdot r_0^2 \cdot R_0 \, d\theta \cdot (1 + \frac{\frac{R_0}{R_0 + a} \cdot \sin\theta}{\sqrt{1 - (\frac{R_0}{R_0 + a})^2 \cdot \cos^2\theta}})}{(R_0 \sin\theta + (R_0 + a)\sqrt{1 - (\frac{R_0}{R_0 + a})^2 \cdot \cos^2\theta})^2} \times (1 - \frac{V_I^2}{c^2} + \frac{3}{2} \cdot \frac{V_I^2}{c^2} \cdot \sin^2\theta) \cdot \sqrt{1 - (\frac{R_0}{R_0 + a})^2 \cdot \cos^2\theta} = \quad (5)$$

$$= (\frac{R_0}{R_0 + a} = \alpha_+) = \frac{-|e| \cdot \overline{n} \cdot \pi \cdot r_0^2}{R_0} \times \frac{\alpha_+^2 \cdot (1 - \frac{V_I^2}{c^2} + \frac{3}{2} \cdot \frac{V_I^2}{c^2} \cdot \sin^2\theta)}{\alpha_+ \sin\theta + \sqrt{1 - \alpha_+^2 \cdot \cos^2\theta}} \cdot d\theta$$

For an almost stationary ring of ions, the expression for $E_r(a)_{+}$ will be similar to (5), but with a + sign and with $V_I = 0$.

The total electric field at the point under study will be:

$$E_r(a) \cong \frac{|e| \cdot \overline{n} \cdot \pi \cdot r_0^2}{R_0} \cdot \frac{V_I^2}{c^2} \cdot 2 \cdot \int_{-\frac{\pi}{2}}^{\frac{\pi}{2}} \frac{\alpha_+^2 \cdot (1 - \frac{3}{2} \cdot \sin^2\theta)}{\alpha_+ \sin\theta + \sqrt{1 - \alpha_+^2 \cdot \cos^2\theta}} \cdot d\theta =$$

$$= \frac{I_p^2}{|e| \cdot \overline{n} \cdot \pi \cdot r_0^2 \cdot R_0} \cdot \frac{1}{c^2} \cdot 2 \cdot \int_{-\frac{\pi}{2}}^{\frac{\pi}{2}} \frac{\alpha_+^2 \cdot (1 - \frac{3}{2} \cdot \sin^2\theta)}{\alpha_+ \sin\theta + \sqrt{1 - \alpha_+^2 \cdot \cos^2\theta}} \cdot d\theta = \quad (6)$$

$$= \frac{I_p^2}{|e| \cdot \overline{n} \cdot \pi \cdot r_0^2 \cdot R_0} \cdot \frac{1}{c^2} \cdot 2 \cdot F(\alpha_+)$$

It is easy to show that for case b), FIG.2., the radial electric field at the observation point will be:

$$E_r(-a) = \left(\frac{R_0}{R_0 - a} = \alpha_-\right) \cong \frac{|e| \cdot \overline{n} \cdot \pi \cdot r_0^2}{R_0} \cdot \frac{V_l^2}{c^2} \cdot 2 \cdot \left(\int\limits_{\frac{\pi}{2}}^{\theta\min} \frac{\alpha_-^2 \cdot (1 - \frac{3}{2} \cdot \sin^2\theta)}{-\alpha_- \sin\theta + \sqrt{1 - \alpha_-^2 \cdot \cos^2\theta}} \cdot d\theta + \right.$$

$$+ \int\limits_{\theta\min}^{\frac{\pi}{2}} \frac{\alpha_-^2 \cdot (1 - \frac{3}{2} \cdot \sin^2\theta)}{\alpha_- \sin\theta + \sqrt{1 - \alpha_-^2 \cdot \cos^2\theta}} \cdot d\theta) = \frac{I_p^2}{|e| \cdot \overline{n} \cdot \pi \cdot r_0^2 \cdot R_0} \cdot \frac{1}{c^2} \cdot 2 \times$$

$$\times (\int\limits_{\frac{\pi}{2}}^{\theta\min} \frac{\alpha_-^2 \cdot (1 - \frac{3}{2} \cdot \sin^2\theta)}{-\alpha_- \sin\theta + \sqrt{1 - \alpha_-^2 \cdot \cos^2\theta}} \cdot d\theta + \int\limits_{\theta\min}^{\frac{\pi}{2}} \frac{\alpha_-^2 \cdot (1 - \frac{3}{2} \cdot \sin^2\theta)}{\alpha_- \sin\theta + \sqrt{1 - \alpha_-^2 \cdot \cos^2\theta}} \cdot d\theta) =$$

$$= \frac{I_p^2}{|e| \cdot \overline{n} \cdot \pi \cdot r_0^2 \cdot R_0} \cdot \frac{1}{c^2} \cdot 2 \cdot (F_1(\alpha_-) + F_2(\alpha_-))$$

(7),

where $\cos(\theta\min) = \dfrac{1}{\alpha_-}$.

Integral expressions in $F(\alpha_+)$, $F_1(\alpha_-)$ and $F_2(\alpha_-)$ (equations (6) and (7)) are very interesting. They are shown in FIG. 3-5 for points $R_0 + a$ and $R_0 - a$, with a/R$_0$=0.3.

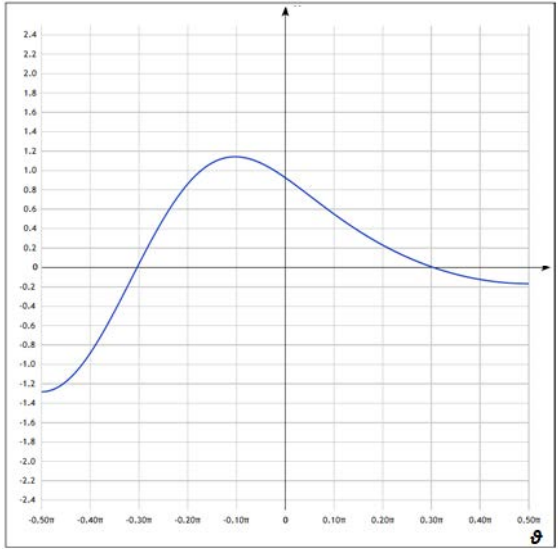

FIG.3. Integrand of $F(\alpha_+)$ in equation (6) for the point $R_0 + a$ (a/R$_0$=0.3). The numerical value of the integral is 0.64.

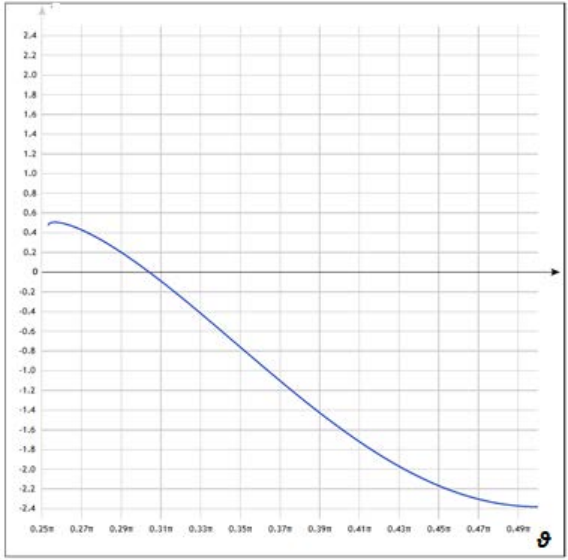

FIG.4. The first part of the integrand in $F_1(\alpha_-)$, see the equation (7), for the point $R_0 - a$ where a/R$_0$=0.3. The numerical value of the integral is -0.85.

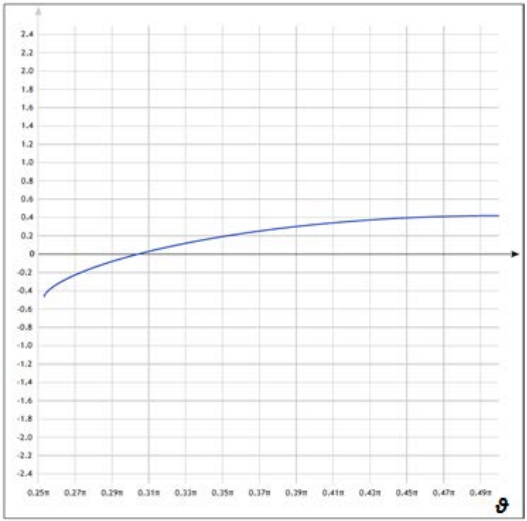

FIG.5. The second part of the integrand in $F_2(\alpha_-)$, see the equation (7), for the point $R_0 - a$ where a/R$_0$=0.3. The numerical value of the integral is 0.15.

Attention is drawn to the fact that, unlike the calculations for FIG.1, the electric field in this case has different signs in $R_0 + a$ ( $E_r(a) \sim +0.64$) and in $R_0 - a$ (total $\sim$ -0.7). Due to the symmetry of the investigating model, the total flow of $E_r(r)$ through a narrow ribbon with a small radius **a** must be zero, according to the Gauss theorem. It is not surprising that the absolute values of the fields on the outer and inner rounds of the torus are close to each other.

Если измеряемая точка с радиусом **a** находится вблизи стенки токамака, в случае заземления камеры только часть плазменного кольца будет определять $E_r(a)$. Легко показать, что для внешнего обхода, формула (6), нужно интегрировать от некого $\theta_{kr}$ до $\frac{\pi}{2}$, где $\sin(\theta_{kr}) = \frac{R_0 - a}{R_0 + a}$. В случае внутреннего обхода тора в формуле (7) остается только первый интеграл. Электрическое поле с учетом заземленной камеры незначительно изменится. $E_r(a)$ станет пропорциональным +0.66 в точке $R_0 + a$, и коэффициент пропорциональности уменьшится до -0.85 в $R_0 - a$.

На РИС. 6 представлены вычисленные интегралы $F_1(\alpha_-) + F_2(\alpha_-)$ для внутреннего обхода тора – отрицательные значения, и $F(\alpha_+)$ для внешнего обхода – положительные значения, для разных величин a/R$_0$.

If the measured point with radius **a** is located near the wall of the tokamak, in the case of grounding the camera, only part of the plasma ring will determine $E_r(a)$. It is easy to show that for an external bypass, equation (6), you need to integrate from some $\theta_{kr}$ to $\frac{\pi}{2}$, where $\sin(\theta_{kr}) = \frac{R_0 - a}{R_0 + a}$. In the case of an internal torus traversal, only the first integral remains in equation (7). The electric field, taking into account the grounded chamber, will change slightly. $E_r(a)$ will become proportional to +0.66 at the point $R_0 + a$, and the proportionality coefficient will decrease to -0.85 at the point $R_0 - a$.

FIG. 6 shows the calculated integrals $F_1(\alpha_-) + F_2(\alpha_-)$ - negative values for the inner circumvent of the torus, and positive values for the outer circumvent $F(\alpha_+)$, for different values a/R$_0$.

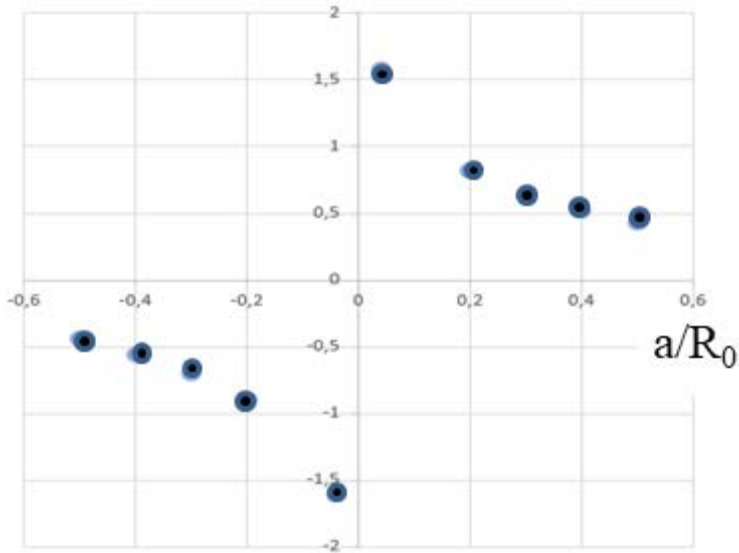

FIG. 6. The dependence of the integrals in equations (6) and (7) on a/$R_0$.

In a real tokamak, the small radius of the plasma is close in magnitude to the small radius of the toroidal chamber. Therefore, equations (6) and (7) can only be used approximately. Calculate the magnitude of the radial electric field for the tokamak T-11M on the external and internal bypass of the chamber. We believe that $r_0 \approx a$. The ratio a/$R_0$ ~ 0.3. The value of integrals in formulas is taken from FIG.6. Then in the GHS system:

$$E_r(a) \cong \frac{I_p^2}{|e| \cdot \bar{n} \cdot \pi \cdot r_0^2 \cdot R_0} \cdot \frac{1}{c^2} \cdot 2 \cdot 0.66 \cong$$
$$\cong \frac{2 \cdot (50 \cdot 10^3 \cdot 3 \cdot 10^9)^2 \cdot 300}{4.8 \cdot 10^{-10} \cdot 10^{13} \cdot 3.14 \cdot 20^2 \cdot 70 \cdot (3 \cdot 10^{10})^2} \cdot 0.66 \cong 35 \cdot 0.66 \tag{8}$$

Thus, on the external bypass of the torus $E_r(a) \cong$ +23 V / cm. A similar calculation for a small torus bypass gives $E_r(-a) \cong$ -30 V/cm. It is taken into account here that the camera on the T-11M is grounded.

On a Torus-Supra tokamak with parameters essential for calculations: plasma current 1 MA, Ro=240 cm, $r_0$=72 cm, $\bar{n} \approx 4 \cdot 10^{13}$ см$^{-3}$, measurements of carbon C$^{+6}$ toroidal rotation velocity on the magnetic surface with a radius of 10 cm from the weak and strong toroidal magnetic fields were carried out [1]. The method of the charge exchange recombination spectroscopy (CXRS) was applied.

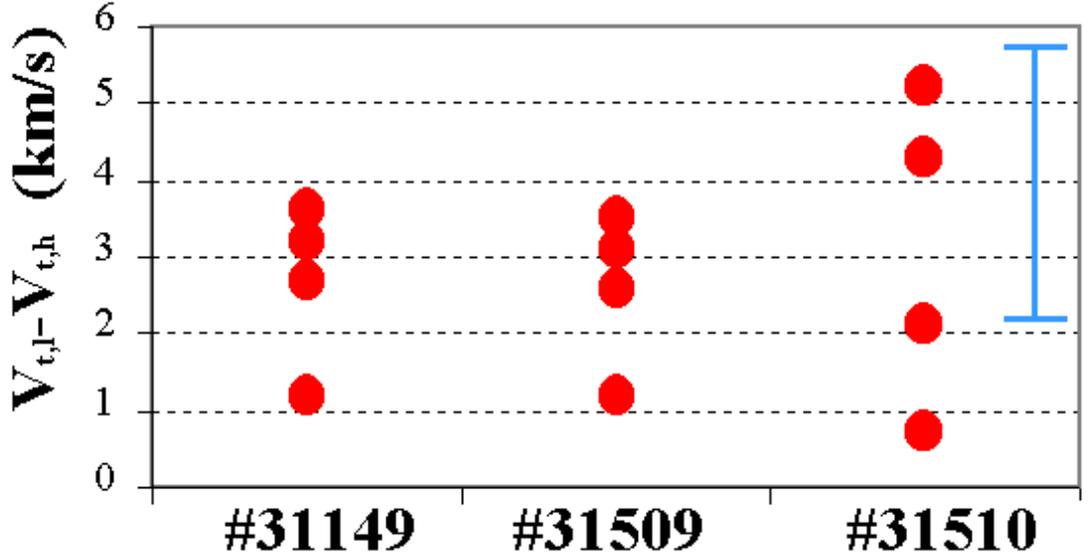

FIG.7. Experiment on measuring the asymmetry of the toroidal rotation velocity $C^{+6}$ for a magnetic surface r≅10 cm in a Tore-Supra tokamak. The vertical axis is the experimental difference of the toroidal velocity $C^{+6}$ between the points on the outer and inner circumferences of the torus in km/s, #31149, #31509, #31510 – the discharge numbers in which the presented velocity measurements were carried out. The accuracy of measuring the velocity difference was approximately 3.5 km/sec.

In FIG.7 presents the measurement results for three discharges from [1]. The difference in the toroidal rotation velocities on the outer and inner circumference of the torus should have been approximately minus 1.5 km/s in the case of plasma rotation as a solid. The plasma rotation velocity for these ohmic discharges was minus 20-25 km/s (counter current direction) [5]. The measured toroidal rotation velocity differences to be close to 3 km/s and the sign of the difference was positive. With a certain accuracy, it is possible to calculate according to the proposed model, considering that the poloidal asymmetry is formed by a plasma inside the radius of the magnetic surface equal to 10 cm. The calculation of the poloidal non-uniformity within the framework of the model under consideration gives: $E_r(+10cm)$ ~2.7 V / cm – on the outer circumference of the torus, and $E_r(-10cm)$ ~ -2.9 V / cm – on the inner circumference of the torus. Based on the model for the radial electric field in the case of large toroidal magnetic field ripples [4, 5], namely such a model is most often considered for the TORUS-SUPRA tokamak, $E_r(r) \approx \frac{1}{|e|}(\frac{T_i}{n_i}\frac{dn_i}{dr} + 3.5\frac{dT_i}{dr})$.

The radial electric field from this formula for a small radius of 10 cm will be approximately minus 28 V/cm. The calculated poloidal non-uniformity of the toroidal rotation of the plasma, equal to about 5 km/s, has a positive sign and, taking into account the toroidal rotation on the magnetic surface as a solid (which would give the non-uniformity associated with the poloidal magnetic field, approximately equal to minus 1.5 km/s), the difference in the toroidal rotation velocities on a given magnetic surface from the external bypass of the torus in relation to the internal bypass is approximately equal to 3.5 km/s. This value correlates well with the experiment.

## Conclusion

A relativistic model of the poloidal non-uniformity radial electric field formation based on: 1) deformations of the electric field of a rapidly moving electron in the longitudinal direction of velocity and perpendicular to it, see equation (1); 2) a simple model of the electric field formation in a toroidal plasma system with a current, see FIG.2. At the same time, a positive additive to $E_r(r)$ appears on the side of the external bypass of the torus, and a negative additive appears on the side of the internal bypass, see equations (6) and (7). These additives are proportional to the square of the plasma current. At the same time, the absolute value of the additives is close to each other, which is not surprising in the framework of the Gauss theorem.

Since the plasma in the tokamak is located in a grounded metal chamber, it is necessary to take into account the effect of shielding on the formation of a poloidal non-uniformity of the radial electric field.

The non-uniformity of the radial electric field considered in this model can lead to a significant non-uniformity of the plasma toroidal rotation on magnetic surfaces, which differs significantly from the traditional representation when the plasma rotates as a solid. At the same time, on the inner bypass of the torus, the plasma will rotate much faster than on the outer bypass of the torus in the case of counter current plasma rotation. A similar result was observed in experiments on the TORUS-SUPRA tokamak [1]. The calculations carried out within the framework of the model under consideration correlate well with these experiments. The mechanism presented in this paper can partially explain the physics of the radial electric field formation, especially at the initial stage of plasma formation and in experiments with a sharp controlled change in plasma current.